 \documentclass[final,5p,times,twocolumn]{elsarticle}

\usepackage{amssymb}
\usepackage{lipsum}
\usepackage{bm}
\usepackage{amsmath}
\usepackage{booktabs}

\usepackage[bookmarks,bookmarksnumbered]{hyperref}
\hypersetup{colorlinks = true,
            linkcolor = blue,
            anchorcolor =red,
            citecolor = blue,
            pdfauthor=author}
\usepackage{float}

\journal{Journal of Subatomic Particles and Cosmology}

\begin{document}

\begin{frontmatter}

\title{Deep learning for exploring hadron-hadron interactions}

\author[first]{Lingxiao Wang}
\affiliation[first]{organization={Interdisciplinary Theoretical and Mathematical Sciences Program (iTHEMS)},%Department and Organization
            addressline={RIKEN}, 
            city={Wako},
            postcode={351-0198}, 
            state={Saitama},
            country={Japan}}

\begin{abstract}

In this proceeding, we introduce deep learning technologies for studying hadron-hadron interactions. To extract parameterized hadron interaction potentials from collision experiments, we employ a supervised learning approach using Femtoscopy data. The deep neural networks (DNNs) are trained to learn the inverse mapping from observations to potentials. To link between experiments and first-principles simulations, we further investigate hadronic interactions in Lattice QCD simulations from the HAL QCD method perspective. Using an unsupervised learning approach, we construct a model-free potential function with symmetric DNNs, aiming to learn hadron interactions directly from simulated correlation functions (equal-time Nambu-Bethe-Salpeter amplitudes). On both fronts, deep learning methods show great promise in advancing our understanding of hadron interactions.

\end{abstract}

%%Graphical abstract
%\begin{graphicalabstract}
%\includegraphics{grabs}
%\end{graphicalabstract}

%%Research highlights
%\begin{highlights}
%\item Research highlight 1
%\item Research highlight 2
%\end{highlights}

\begin{keyword}
%% keywords here, in the form: keyword \sep keyword, up to a maximum of 6 keywords
Deep Learning \sep Hadron Physics \sep Femtoscopy \sep Lattice QCD

%% PACS codes here, in the form: \PACS code \sep code

%% MSC codes here, in the form: \MSC code \sep code
%% or \MSC[2008] code \sep code (2000 is the default)

\end{keyword}

\end{frontmatter}

\section{Introduction}
\label{sec:intro}

Hadron-hadron interactions describe the fundamental dynamics between two or more hadrons, such as protons and neutrons, or other baryons and mesons. These interactions are a manifestation of the strong interaction, which binds quarks together inside hadrons and extends to produce forces between hadrons. The earliest attempt to explain nuclear forces was made by H. Yukawa~\cite{Yukawa:1935xg}, proposing that the forces between nuclei are mediated by the exchange of mesons, notably the $\pi$ meson, leading to the development of the Yukawa potential. Later discoveries of heavier mesons expanded it into the highly successful one-boson-exchange (OBE) framework~\cite{Bryan:1964zzb,PhysRev.170.907}, which remains a cornerstone for describing interactions between hadrons.

Modern theories view hadron-hadron interactions as a residual effect of Quantum Chromodynamics (QCD), the fundamental theory of the strong interaction. Based on QCD principles, chiral effective field theory has been developed to provide a systematic framework for studying hadron-hadron interactions, incorporating QCD symmetries and allowing the study of interactions involving multiple hadrons~\cite{Entem:2003ft,Machleidt:2011zz,Meng:2022ozq}.

First-principles calculations, such as those using lattice QCD (LQCD), have also been applied to explore hadron-hadron interactions in unprecedented detail~\cite{Wagman:2017tmp,Aoki:2020bew}. In these calculations, spatial correlations between two hadrons are analyzed to extract interaction potentials, with the HAL QCD method providing a robust tool for deriving hadron-hadron forces~\cite{Aoki:2011ep,HALQCD:2018gyl,Hatsuda:2018nes,Aoki:2020bew}.

On the experimental side, direct measurement of hadron interactions is not feasible. A practical and effective approach is Femtoscopy, inspired by the Hanbury Brown and Twiss (HBT) correlation~\cite{Pratt:1984su,Lisa:2005dd}. The Femtoscopy technique connects the correlations of particle pairs produced in high-energy collisions to hadron-hadron interactions at the femtometer scale. Advanced collision experiments, such as ALICE at CERN-LHC and STAR at BNL-RHIC, employ Femtoscopy to measure two-body correlations and probe hadron interactions~\cite{Ohnishi:2023jlx,Fabbietti:2020bfg,ALICE:2020mfd,STAR:2018uho}. Furthermore, this technique has been instrumental in exploring exotic hadronic states~\cite{Liu:2024uxn}.

As a matter of fact, extracting hadron-hadron interaction potentials from Femtoscopy and LQCD is a typical inverse problem in QCD physics. In such problems~\cite{Zhou:2023pti}, the goal is to learn the underlying physical quantity, i.e., interaction potentials, from observable data like particle correlations or equal-time Nambu-Bethe-Salpeter(NBS) amplitudes. Deep learning~\cite{lecun2015deep}, as a modern and the most successful branch of machine learning, offers a complementary solution for inverse problems~\cite{tanaka2021deep}, with its capacity to model complex and non-linear relationships using deep neural networks (DNNs). The flexibility of deep learning frameworks enables them to generalize across diverse datasets and account for subtleties inherent in hadron-hadron systems, providing a robust tool for solving inverse problems in QCD physics.

\section{Learning from Femtoscopy}
\label{sec:fem}

\subsection{Femtoscopy}
Based on the femtoscopic formalism, the correlation function can be computed by convolving the source function $S({\bm r})$ with the two-body scattering wave function $\psi_k({\bm r})$ as follows,
\begin{eqnarray}
\label{eq.correlation}
C(k) = \int S({\bm r}) |\psi_k({\bm r})|^2 d{\bm r},
\end{eqnarray}
where $k = |{\bm p}_1 - {\bm p}_2|/2$ represents the relative momentum. The two-body scattering wave function, $\psi_k({\bm r})$, can be determined by solving the two-body Schr\"odinger equation. Under the approximation of a short-range interaction potential, the wave function can be expressed in terms of the scattering length, $a_0$, and the effective range of the potential, $r_{\rm eff}$. In this scenario, with a Gaussian-shaped source function, an analytical solution for $C(k)$ exists, as first developed by Lednick\'y and Lyuboshits~\cite{Lednicky:1981su}. 

To achieve a more precise understanding of hadron interactions, a general framework is required for reconstructing correlation measurements. The numerical evaluation of the full wave function via the Schrödinger equation, combined with the computation of $C(k)$ for various source functions, has been implemented in two widely used tools: the Correlation Afterburner (CRAB)\cite{crabcite} and the “Correlation Analysis Tool using the Schrödinger equation” (CATS)\cite{Mihaylov:2018rva,Fabbietti:2020bfg}. With an estimated source function~\cite{Wang:2024bpl}, the correlation function $C(k)$ provides a correspondence to the interaction potential. Consequently, using numerous experimentally observed correlation functions, the hadron interaction potentials can be extracted inversely with the assistance of deep learning techniques~\cite{Zhou:2023pti}.

\subsection{Data Preparation}
Here we employ CATS to generate training datasets. CATS provides a ﬂexible tool to calculate correlation function $C(k)$ for different sources and interaction potentials. In this section, we utilize a parameterized form of the hadron-hadron potential and a Gaussian source. The parameters of the potential and the radius of the source are randomly sampled within reasonable ranges. Using CATS, we calculate $C(k)$ for different potentials and sources, which are then used as labelled data. 

%%%%%%%%%%%%%%%%%%%%%%%%%%%%%%%%%%%%%%%%%%%%%%%%%%%%%%%%
\begin{figure}[!htbp]
\begin{center}
\includegraphics[width=0.45\textwidth]{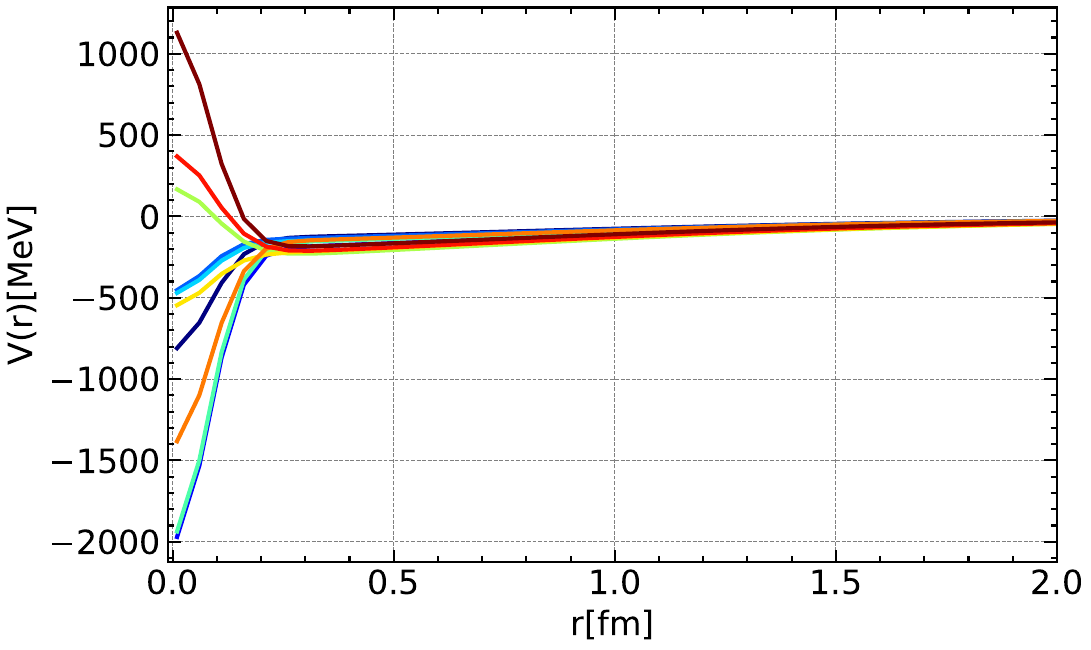}
\caption{Hadron-hadron interaction potentials with multi-Gaussian functions as parameterizations in 10 samples.}
\label{fig:pot}
\end{center}
\end{figure}
%%%%%%%%%%%%%%%%%%%%%%%%%%%%%%%%%%%%%%%%%%%%%%%%%%%%%%%%

\begin{figure}[!htbp]
\begin{center}
\includegraphics[width=0.45\textwidth]{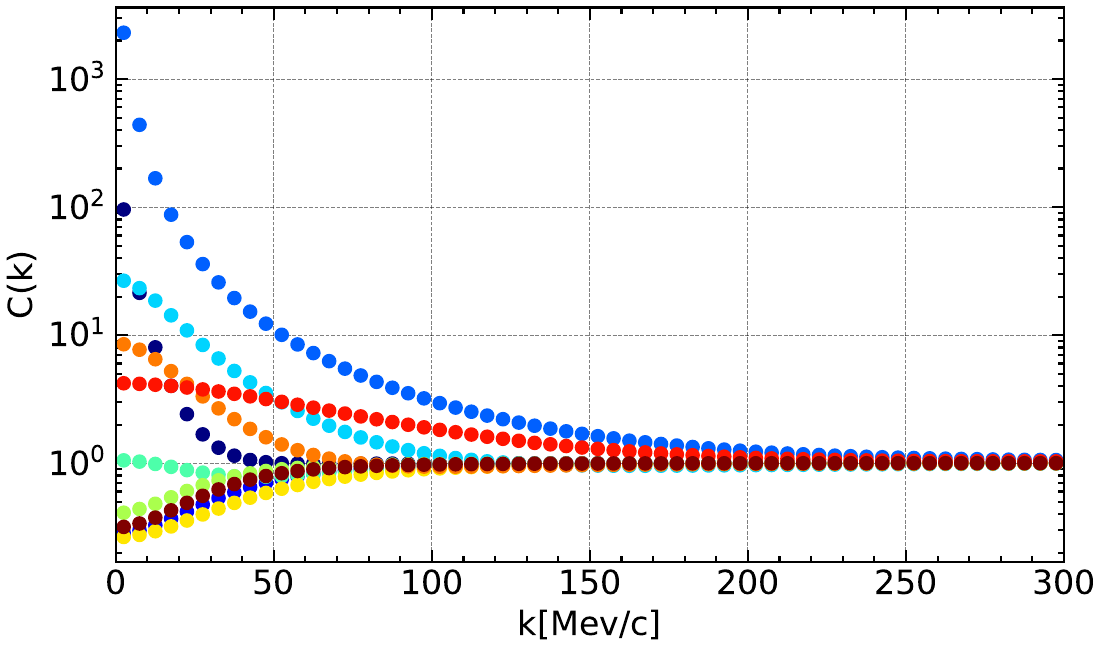}
\caption{CATS computed correlation functions from the paramterized potential functions.}
\label{fig:corr}
\end{center}
\end{figure}

The following parameterization is used to model the hadron interaction potential function,
\begin{equation}
\label{parameterized_pot}
V(r) = b_1 e^{-b_2 r^2} + b_3 (1 - e^{-b_4 r^2}) \left(\frac{e^{-m_{\pi} r}}{r}\right)^2,
\end{equation}
as introduced in~\cite{Aoki:2012tk}. The parameters $b_1$, $b_2$, $b_3$, and $b_4$ have physical units of $\text{MeV}$, $\text{fm}^{-2}$, $\text{MeV} \cdot \text{fm}^2$, and $\text{fm}^{-2}$, respectively, with the pion mass given by $m_{\pi} \approx 135\ \text{MeV}$. For simplicity, spin-dependent interactions and other fine effects are neglected in this initial analysis. By varying the potential parameters within reasonable ranges, a variety of potentials can be generated, encompassing both repulsive and attractive interactions. The parameter ranges are as follows: $b_1$ varies within $[-2145.5, 1532.5]$, $b_2$ spans $[44.34, 103.46]$, $b_3$ lies within $[-345.8, -186.2]$, and $b_4$ ranges from $[0.702, 0.858]$. These ranges enable the exploration of diverse interaction scenarios, and 10 samples are demonstrated in Figure~\ref{fig:pot}, the corresponding correlation functions can be found in in Figure~\ref{fig:corr} with same colors.

%%%%%%%%%%%%%%%%%%%%%%%%%%%%%%%%%%%%%%%%%%%%%%%%%%%%%%%%
%%%%%%%%%%%%%%%%%%%%%%%%%%%%%%%%%%%%%%%%%%%%%%%%%%%%%%%%

In the computation of correlation functions, we employ a Gaussian source function, where the radius $r_0$ serves as the controlling parameter for the source size, as described by,
\begin{equation}
\label{eq:source}
S(r) = \frac{1}{(4\pi r_0^2)^{3/2}} \exp\left(-\frac{r^2}{4r_0^2}\right).
\end{equation}
While determining the precise shape of source functions in different collision systems is non-trivial, we focus on the Gaussian-type source with $r_0$ varying in the range $[0.52, 4.16]\ \text{fm}$. Notably, as shown in Ref.~\cite{Wang:2024bpl}, one can construct a more accurate source function using deep neural networks when the potential functions are known.
%%%%%%%%%%%%%%%%%%%%%%%%%%%%%%%%%%%%%%%%%%%%%%%%%%%%%%%%
\begin{figure}[!htbp]
\begin{center}
\includegraphics[width=0.45\textwidth]{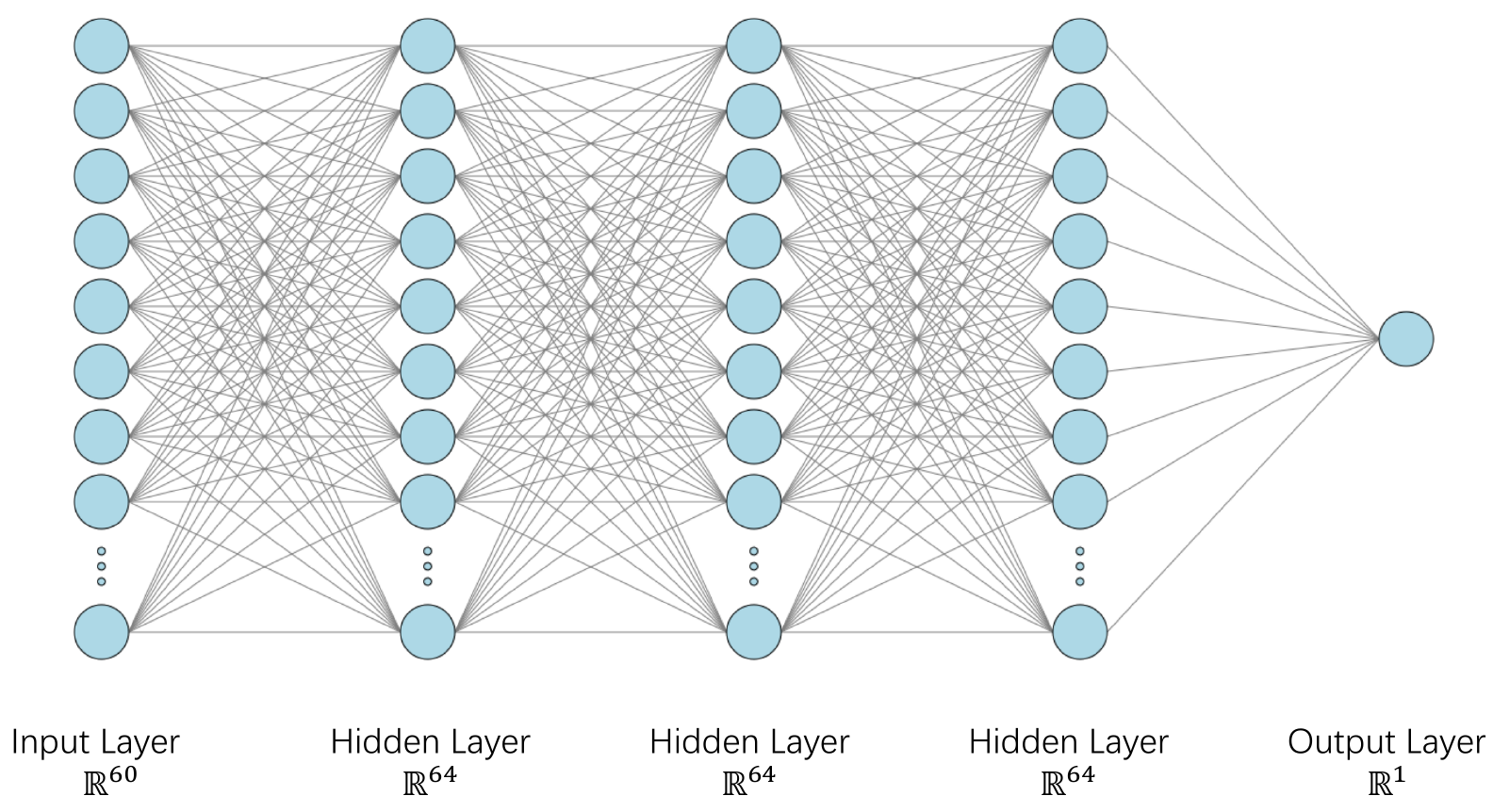}
\caption{The deep neural network(DNN) for predicting the target parameters in potential functions from input correlation functions.}
\label{fig:dnn}
\end{center}
\end{figure}
%%%%%%%%%%%%%%%%%%%%%%%%%%%%%%%%%%%%%%%%%%%%%%%%%%%%%%%%

In the data preparation process, we divided the regression task into two distinct scenarios. The first scenario involves predicting one of the parameters from correlation functions computed using CATS, given 60 $k$-measurements uniformly distributed from $0$ to $300\ \text{MeV}/c$. For this case, we prepared 6,400 samples for each parameter, including the source size. Unless otherwise specified, the fixed parameter values are $r_0 = 1.3\ \text{fm}$, $b_1 = -306.5\ \text{MeV}$, $b_2 = 73.9\ \text{fm}^{-2}$, $b_3 = -266\ \text{MeV} \cdot \text{fm}^2$, and $b_4 = 0.78\ \text{fm}^{-2}$. The parameter being predicted is uniformly sampled from the ranges described earlier. In the second scenario, we aim to predict $b_1$ and $b_3$ simultaneously, using 10,000 samples. The remaining setup is identical to the first scenario. These two parameters were selected because they are relatively more influential in determining the shape of the potential functions. In both scenarios, the dataset is split into training and testing sets with a ratio of $8:2$.

\subsection{Learning Inverse Mapping}

We develop a deep neural network (DNN) model to learn the inverse mapping from correlation functions to the parameters of potential functions. As shown in Figure~\ref{fig:dnn}, the basic architecture of the DNN consists of one input layer, three hidden layers, and one output layer. The input layer contains 60 nodes, corresponding to the required size of the correlation function input. Each hidden layer consists of 64 nodes. All non-activation functions between layers are chosen as ELU~\cite{clevert2015fast}.The final output layer is designed to predict the target label, which, in the simplest case, corresponds to one of the potential parameters.

%%%%%%%%%%%%%%%%%%%%%%%%%%%%%%%%%%%%%%%%%%%%%%%%%%%%%%%%
\begin{table}[!hbpt]
\centering
\caption{$R^2$ metric for testing data sets}
\label{tab:eva}
\begin{tabular}{@{}cccccc@{}}
\toprule
        & $r_0$ & $b_1$ & $b_2$ & $b_3$ & $b_4$ \\ \midrule
Model-1 & 0.998  & 0.999  &0.998    &0.999    &0.998   \\
Model-2 & -  & 0.998  & -  & 0.998 & -  \\ \bottomrule
\end{tabular}
\end{table}
%%%%%%%%%%%%%%%%%%%%%%%%%%%%%%%%%%%%%%%%%%%%%%%%%%%%%%%%

In the testing set, we use the $R^2$ metric to evaluate the performance of the predictions. The $R^2$ value, or coefficient of determination, quantifies how well the model explains the variance in the target variable. It is defined as,
\begin{equation}
R^2 = 1 - \frac{\sum_{i=1}^{n}(y_i - \hat{y}_i)^2}{\sum_{i=1}^{n}(y_i - \bar{y})^2},
\end{equation}
where $y_i$ are the target values, $\hat{y}_i$ are the predicted values, $\bar{y}$ is the mean of $y_i$, and $n$ is the number of data points. An $R^2$ value of 1 indicates perfect prediction, 0 implies that the model performs no better than predicting the mean, and negative values suggest the model performs worse than the mean prediction. In Table~\ref{tab:eva}, it shows the performance of two models in two scenarios, where Model-1 successfully predicts each parameter, and Model-2 also achieves $R^2=0.998$ for two parameters simultaneously in the testing set.

\section{Learning from LQCD}
\label{sec:LQCD}

\subsection{HAL QCD Method}

The HAL QCD method~\cite{Ishii:2006ec,Aoki:2008hh,Aoki:2009ji,Ishii:2012ssm} has been proposed to build effective potentials between hadrons from their spatial correlations, the equal-time Nambu-Bethe-Salpeter (NBS) amplitude $\phi_{\bf k}({\bf r})$, measured on the lattice, bridging the gap between LQCD and experimental data (see e.g.~\cite{ALICE:2020mfd}). Comprehensive reviews are available in Refs.~\cite{Aoki:2020bew,Aoki:2023qih}. In this method, the integral kernel of the integro-differential equation for the NBS wave function is treated as a non-local potential between hadrons. The non-local potential $U({\bf r},{\bf r}^{\prime})$ 
for two baryons with an equal mass $m_B$ can be defined as ~\cite{Ishii:2006ec,Aoki:2008hh,Aoki:2009ji},  
\begin{eqnarray}
    ({ E}_{\bf k}-{ H}_{0}){\phi}_{\bf k}({\bf r})=\int d^{3}\,r^{\prime}{ U}({\bf r},{\bf r}^{\prime}){\phi}_{\bf k}({\bf r}^{\prime}),\nonumber \\
    \quad{ E}_{\bf k}=\frac{{\bf k}^2}{2m},\quad{H}_{0}=-{\frac{{\nabla}^{2}}{2m}},\quad m={\frac{m_{B}}{2}}.\label{eq:schor}
\end{eqnarray}

Since all the elastic scattering states are governed by the same potential $U({\bf r},{\bf r}^{\prime})$, the time-dependent HAL QCD method~\cite{Ishii:2012ssm} takes full advantage of all the NBS amplitudes below the inelastic threshold  $\Delta E^* \sim \Lambda_{\rm QCD}$ by defining  so-called the $R$ correlator as $R(t, {\bf r})= \sum_{n}^{\infty}A_{n}\psi_{n}({\bf r})e^{-(\Delta W_{n})t}+O(e^{-(\Delta E^{*})t}),$ where $A_n$ is the overlapping factor, and $\Delta W_n = 2\sqrt{m^2_B + {\bf k}_n^2} - 2m_B$ with the relative momentum ${\bf k}_n$. The contributions from the inelastic states are exponentially suppressed when $t\gg (\Delta E^*)^{-1}$. In such condition, the $R$ correlator can be shown to satisfy following integro-differential equation~\cite{Ishii:2012ssm} as follows,
\begin{equation}
    \left\{\frac{1}{4m_{B}}\frac{\partial^{2}}{\partial t^{2}}-\frac{\partial}{\partial t}-H_{0}\right\}R(t,{\bf r})=\int d^{3}{\bf r^{\prime}}\,U({\bf r},{\bf r}^{\prime})R(t,{\bf r}^{\prime}).\label{eq:timeeq}
\end{equation}
The effective central potential in the leading order approximation of the velocity expansion, $U({\bf r},{\bf r}^{\prime}) = V(r)\delta({\bf r}-{\bf r^{\prime}})+\Sigma_{n=1}V_{2n}({\bf r})\nabla^{2n}({\bf r}-{\bf r^{\prime}})$, can be computed directly from,
\begin{equation}
   V(r) = \frac{1}{R(t,{\bf r})}\left\{\frac{1}{4m_{B}}\frac{\partial^{2}}{\partial t^{2}}-\frac{\partial}{\partial t}-H_{0}\right\}R(t,{\bf r}).
\end{equation}
Also, the higher-order terms of the velocity expansion $V_{2n}$ can be obtained by combining the information of the $R$ correlators obtained from different source operators or equivalently different weight factors $A_n$~\cite{HALQCD:2018gyl}.

\subsection{Neural Network Hadron Potentials}
\begin{figure}[!hbpt]
    \centering
    \includegraphics[width = 0.45 \textwidth]{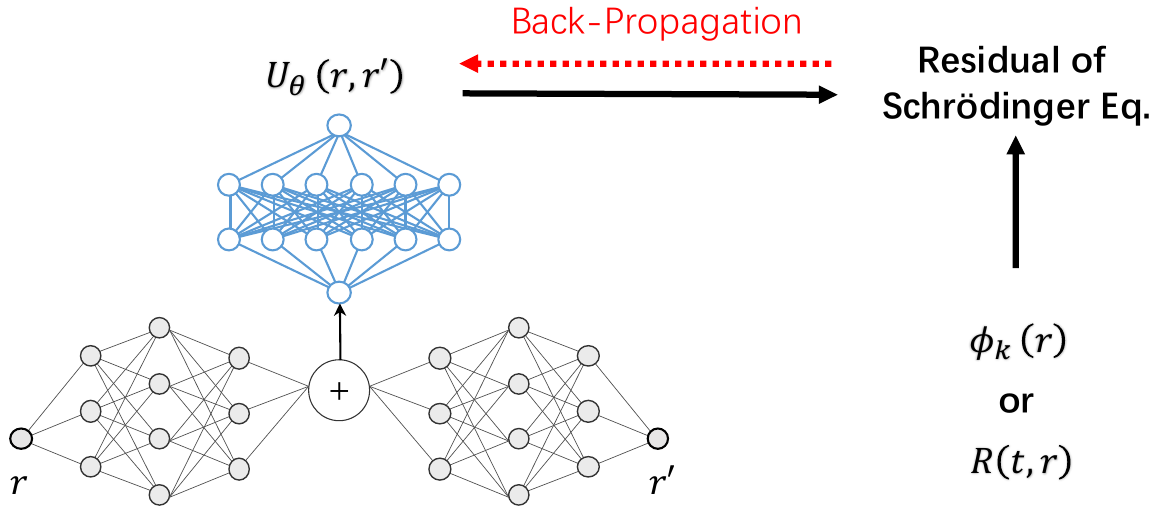}
    \caption{Symmetric deep neural network (SDNN) for representing potential functions. The gray colored neural networks are the same for inputs $r$ and $r^\prime$. The outputs of two different inputs are added in the latent layer, which is used as the input of the next layer neural network colored as blue. The final output represents $U_\theta(r, r^\prime)$.}
    \label{fig:symnn}
\end{figure}

To simplify the reconstruction task without loss of generality, we begin by considering the wave function in the S-wave for systems of two identical particles. To preserve the exchange symmetry inherent in the non-local potential of such systems, we design a parameter-sharing neural network~\cite{Wang:2024ykk}. Figure~\ref{fig:symnn} illustrates a schematic of the symmetric deep neural network (SDNN) used to represent the potential $U_\theta(r, r^\prime)$. The inputs to the network are $(r, r^\prime)$, and the output from the parameter-sharing network is $f(r)$. This output is then combined with $f(r^\prime)$ as input to the subsequent layer. The final output of the network is defined as  $U_\theta(r, r^\prime) \equiv g(f(r) + f(r^\prime))$ , where $g(x)$ and $f(x)$ are two distinct neural networks, and  $\theta$  denotes all the trainable parameters within the neural network.

%%%%%%%%%%%%%%%%%%%%%%%%%%%%%%%%%%%%%%%%%%%%%%%%%%%%%%%%%%%%%%%
%%%%%%%%%%%%%%%%%%%%%%%%%%%%%%%%%%%%%%%%%%%%%%%%%%%%%%%%%%%%%%%
Given the wave function $\phi_{k}({r})$ and the energy $E_k$, the potential $U_\theta({r},{r}^{\prime})$ can be determined by minimizing the loss function as the residual of Eq.~\eqref{eq:schor}. Furthermore, given the correlation function $R(t,r)$, the potential $U_\theta(r ,r')$ can be determined through minimizing the residual of Eq.~\eqref{eq:timeeq}.

To introduce the physical constraint as a regularization, we adopt the asymptotic behaviour of the hadron-hadron interaction, $\lim_{r,r\rightarrow \infty} U(r,r^{\prime}) = 0$, as the regularization loss function,
\begin{equation}
    \mathcal{L}_r =  \sum_{n,m}^N U_\theta(r_n,r^\prime_m) ^2, \quad r_n>\tilde{R}, \quad  r^{\prime}_m >\tilde{R},
\end{equation}
where $\tilde{R}$ is a cutoff for indicating there is zero potential. The total loss function becomes, $\mathcal{L}\equiv \mathcal{L}_{\text{data}} + \mathcal{L}_r$. As Figure~\ref{fig:symnn} shows, the wave function $\phi_{k}({r})$ (correlation function $R(t,r)$) and potential function $U_\theta(r ,r')$ are used to compute the residual, and further used to calculate the gradients to parameters of neural networks. The gradient-based algorithm, back-propagation (BP) method~\cite{bishop2023deep}, is applied to optimize the neural network parameters $\{ \theta \}$ by,
\begin{equation}
    \theta_{i+1} \rightarrow \theta_i + \frac{\partial \mathcal{L}}{\partial U_\theta(r,r^\prime)} \frac{\partial U_\theta(r,r^\prime)}{\partial \theta},
\end{equation}
where the index $i$ labels the time-step in optimization process~\cite{Wang:2021jou,Shi:2022yqw}.

\subsection{Separable Potential}
We start from a solvable potential, the separable potential~\cite{Aoki:2020bew} used as a toy model for demonstration. The definition of the radial potential is,
\begin{equation}
    U({\bf r},{\bf r}^{\prime})\,\equiv\,\omega\nu({\bf r})\nu({\bf r}^{\prime}),\;\;\;\;\;\nu({\bf r})\equiv e^{-\mu r},
\end{equation}
where $\omega, \mu$ are parameters. The S-wave solution of the Schr\"odinger equation with this potential is given exactly by,
\begin{equation}
    {{\phi_{k}^{0}(r)}} {{=\displaystyle\frac{e^{i\delta_{0}(k)}}{k r}\left[\sin(k r+\delta_{0}(k))-\sin\delta_{0}(k)e^{-\mu r}\left(1+\displaystyle\frac{r(\mu^{2}+k^{2})}{2\mu}\right)\right]}}.
\end{equation}
As a numerical example, we take $\mu=1.0$, $\omega=-0.017 \mu^4$ and $m=3.30 \mu$ and $R = 2.5/\mu$, the physics unit is chosen as $\mu$. 

When setting ${\varphi}^0_{ k}({ r}) / r\equiv {\phi}^0_{ k}({ r})$, in the spherically symmetric case, the radial euqation will be derived as,
\begin{equation}
    \left( \frac{d^2}{dr^2} + k^2 \right) {\varphi}^0_{ k}(r) = 8\pi mr \int r^{\prime} dr^{\prime} U(r, r^{\prime}) {\varphi}^0_{ k}(r^{\prime}).
\end{equation}

%%%%%%%%%%%%%%%%%%%%%%%%%%%%%%%%%%%%%%%%%%%%%%%%%%%%%%%%%%%%%%%%%%%%%%%%
\begin{figure}[!hbpt]
    \centering
    \includegraphics[width=0.5\textwidth]{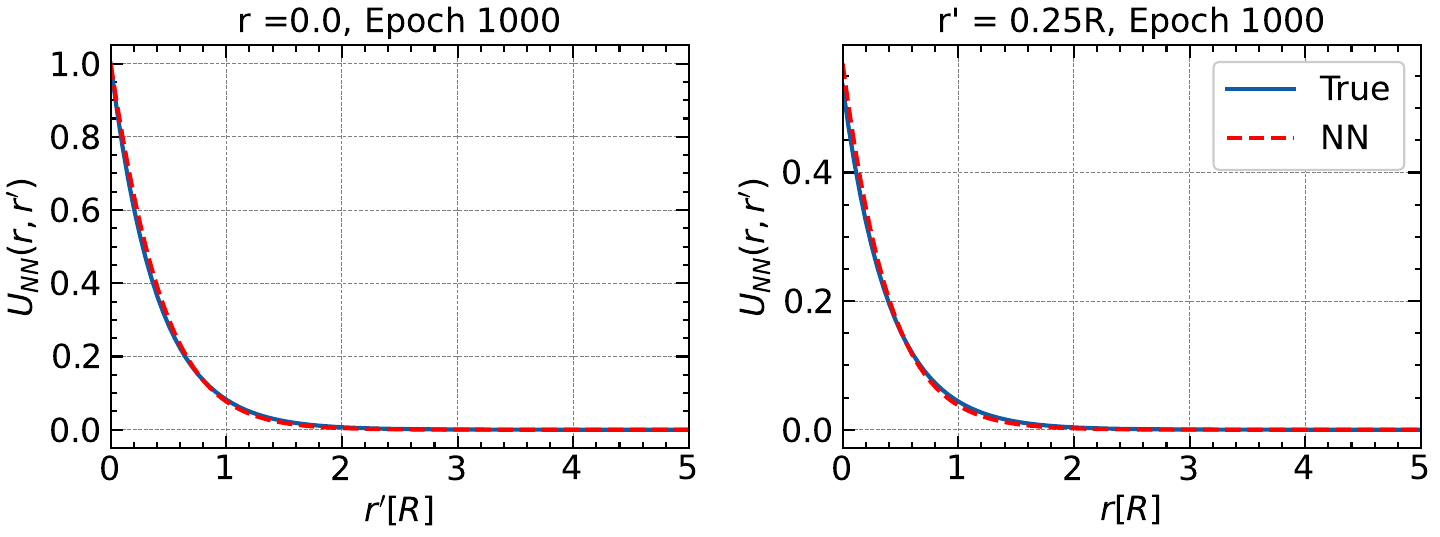}
    \caption{Reconstructed separable potentials from neutral network (NN) and the ground truths.}
    \label{fig:pot-sep}
\end{figure}
%%%%%%%%%%%%%%%%%%%%%%%%%%%%%%%%%%%%%%%%%%%%%%%%%%%%%%%%%%%%%%%%%%%%%%%%

A practical setup for preparing wave functions involves using momentum values $k = [0.01, 1.0]$ with $N_k = 10$ and radial distances $r = [0.01, 5R]$ with $N_r = 100$. A total of 1000 data points are employed to minimize the loss function for an optimal potential $U_{NN}(r, r^\prime) \equiv \omega U_\theta(r, r^\prime)$. The SDNN configuration, as illustrated in Fig.~\ref{fig:symnn}, consists of two identical network paths that process inputs $r$ and $r^\prime$ through a series of linear transformations (structured as $1 \rightarrow 64 \rightarrow 16$) with LeakyReLU activations. These outputs are additively combined to enforce symmetry, and the merged feature is passed through a final linear layer (structured as $16 \rightarrow 1$). A Softplus activation function is applied to the output to ensure smooth, positive predictions. The reconstructed potential is shown in Fig.~\ref{fig:pot-sep}. Regularization is achieved by imposing the asymptotic behavior $U_\theta(r > \tilde{R}, r^\prime > \tilde{R}) = 0$, where $\tilde{R} = 2R$. After 1000 epochs of training, the symmetric deep neural network successfully recovers the ground truth potential functions.

\subsection{Charm Hadron Interactions}

%%%%%%%%%%%%%%%%%%%%%%%%%%%%%%%%%%%%%%%%%%%%%%%%%%%%%%%%%%%%%%%%%%%%%%%%
\begin{figure}[!hbpt]
    \centering
    \includegraphics[width=0.42\textwidth]{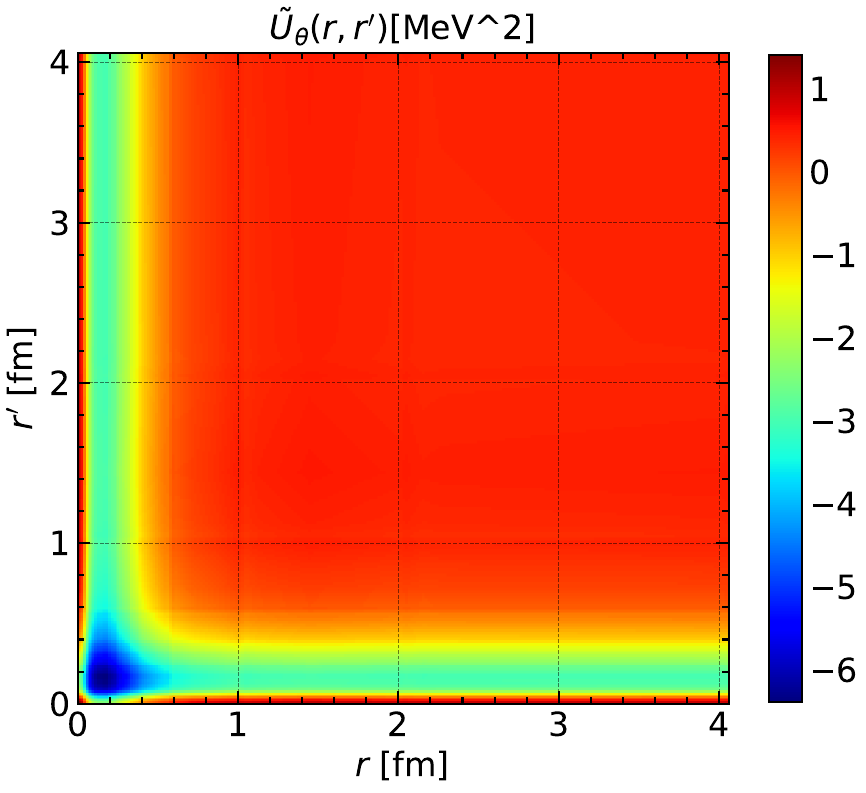}
    \caption{Non-local neural network potential for $\Omega_{ccc}-\Omega_{ccc}$.
    }
    \label{fig:pot-oo-2}
\end{figure}
%%%%%%%%%%%%%%%%%%%%%%%%%%%%%%%%%%%%%%%%%%%%%%%%%%%%%%%%%%%%%%%%%%%%%%%%

For the second demonstration, we focus on the $\Omega_{ccc}-\Omega_{ccc}$ system, which has been extensively studied in recent research~\cite{Lyu:2021cte,Lyu:2021qsh}. The gauge configurations employed in this work utilize a $(2+1)$-flavor setup on a $96^4$ lattice with the Iwasaki gauge action at $\beta = 1.82$. The lattice spacing is approximately $a \approx 0.0846 , \text{fm}$ ($a^{-1} \approx 2.333 , \text{GeV}$), with the pion and kaon masses set to $m_\pi \approx 146 , \text{MeV}$ and $m_K \approx 525 , \text{MeV}$, respectively. The interpolated mass of the $\Omega_{ccc}$ baryon is $m_{\Omega_{ccc}} \approx 4796 , \text{MeV}$. Further details regarding the lattice setup can be found in Ref.~\cite{Lyu:2021qsh}.

In this setup, the symmetric deep neural network (SDNN) architecture is deeper than in the previous case, following the structure ($1 \rightarrow 32 \rightarrow 64 \rightarrow 128 \rightarrow 64 \rightarrow 32 \rightarrow 16$), with an ELU activation applied to the final output. Other configurations remain identical to those in the previous case. The model was trained using $R(t = 26, r)$ correlation data of the $\Omega_{ccc}-\Omega_{ccc}$ system calculated from lattice QCD simulations, where $t = 25$ and $t = 27$ were used exclusively for computing $R_{tt}$ and $R_t$. Regularization was applied by enforcing the asymptotic condition $U_\theta(r > 3,\text{fm}, r’ > 3,\text{fm}) = 0$. After 5000 epochs, the non-local 3D potential function, defined as $\tilde{U}_\theta(r, r’) \equiv 4\pi {r^{\prime}}^2 {U}_\theta(r, r’)$ with $\Delta r’ = 0.2a$, is demonstrated for the first time in Fig.~\ref{fig:pot-oo-2}.

\section{Conclusions}

\subsection{Summary}
This proceeding introduces deep learning to investigate hadron-hadron interactions. In the supervised learning approach, experimental-type Femtoscopy data are mapped to interaction potentials, while unsupervised learning, employing symmetric neural networks, derives potentials directly from Lattice QCD simulations. These methodologies demonstrate significant potential for bridging experimental observations with theoretical models, offering new insights into hadron interactions and advancing our understanding of their underlying dynamics.

\subsection{Outlooks}
In supervised learning, reliable simulation data sets must first be prepared, yielding physically interpretable but potentially biased results. An attempt to predict all five parameters simultaneously, however, inevitably leads to failure. In our preliminary results, this is supported by the principal component analysis (PCA)~\cite{greenacre2022principal}, which revealed that at least the first two or three components are sufficient to explain the variance in the correlation function measurements. This finding highlights the need to identify improved observables beyond correlations for extracting hadron interactions from collision experiments.

On the other hand, physics-driven designed neural networks provide a robust framework to directly extract physical quantities from observational data with greater precision. But the further comparisons between the theoretical predictions and experimental observations  are still missing, in particular, we should compute the scattering phase shifts and scattering parameters more explicitly in our future works.

\section*{Acknowledgements}
I would like to thank the Organizers of EMMI Workshop at the University of Wrocław - Aspects of Criticality II for the invitation to present this talk. I also thank T.~Doi, T.~Hatsuda, Y.~Lyu, L.~Zhang and J.~Zhao for discussions/material covered in my talk. I thank the members of HAL QCD Collaboration. The lattice QCD measurements have been performed on HOKUSAI supercomputers at RIKEN. This material is supported by Japan Science and Technology Agency (JST) as part of Adopting Sustainable Partnerships for Innovative Research Ecosystem (ASPIRE), Grant Number JPMJAP2318. This material is also supported by JSPS (JP19K03879,JP23H05439) and MEXT (JPMXP1020230411).

\bibliographystyle{elsarticle-num} 
\bibliography{emmi_pro}

\end{document}